\documentclass[12pt,preprint]{aastex}

\newcommand{\msun}{\mbox{$M_\odot$}}
\newcommand{\lsun}{\mbox{$L_\odot$}}
\newcommand{\rsun}{\mbox{$R_\odot$}}

\newcommand{\logg}{\mbox{$\log g$}}

\newcommand{\etal}{\mbox{\rm et al.~}}
\newcommand{\ms}{\mbox{m s$^{-1}$}}
\newcommand{\ks}{\mbox{km s$^{-1}$}}

\newcommand{\mjup}{$M_{\rm JUP}$}

\newcommand{\vsini}{$v \sin i~$}

\shorttitle{A 13 Jupiter-mass Companion in a 1.3 yr
Nearly-circular Orbit of HD 16760}
\shortauthors{Sato et al.}

\begin{document}


\title{A Substellar Companion in a 1.3 yr
Nearly-circular Orbit of HD 16760\altaffilmark{1,2}}


\author{Bun'ei Sato\altaffilmark{3},
Debra A. Fischer\altaffilmark{4},
Shigeru Ida\altaffilmark{5},
Hiroki Harakawa\altaffilmark{5},
Masashi Omiya\altaffilmark{6},
John A. Johnson\altaffilmark{7},
Geoffrey W. Marcy\altaffilmark{8},
Eri Toyota\altaffilmark{9},
Yasunori Hori\altaffilmark{5},
Howard Isaacson\altaffilmark{4},
Andrew W. Howard\altaffilmark{8},
and Kathryn M.G. Peek\altaffilmark{8}}

\email{sato.b.aa@m.titech.ac.jp}


\altaffiltext{1}{Based on data collected at the Subaru Telescope, which is
operated by the National Astronomical Observatory of Japan.}
\altaffiltext{2}{Based on observations obtained at the W. M. Keck Observatory,
which is operated by the University of California and the California Institute of
Technology. Keck time has been granted by NOAO and NASA.}
\altaffiltext{3}{Global Edge Institute, Tokyo Institute of Technology,
2-12-1-S6-6 Ookayama, Meguro-ku, Tokyo 152-8550}
\altaffiltext{4}{Department of Physics \& Astronomy, San Francisco State
University, San Francisco, CA  94132}
\altaffiltext{5}{Department of Earth and Planetary Sciences, Tokyo Institute
of Technology, 2-12-1 Ookayama, Meguro-ku, Tokyo 152-8551}
\altaffiltext{6}{Department of Physics, Tokai University,
1117 Kitakaname, Hiratsuka, Kanagawa 259-1292}
\altaffiltext{7}{Institute for Astronomy, University of Hawaii, Honolulu,
HI 96822}
\altaffiltext{8}{Department of Astronomy, University of California,
Berkeley, CA}
\altaffiltext{9}{Kobe Science Museum, 7-7-6 Minatoshima-Nakamachi, Chuo-ku,
Kobe, Hyogo 650-0046}


\begin{abstract}
We report the detection of a substellar companion orbiting the G5 dwarf
HD 16760 from the N2K sample.
Precise Doppler measurements of the star from Subaru and Keck revealed
a Keplerian velocity variation with a period of 466.47$\pm$0.35 d,
a semiamplitude of 407.71$\pm$0.84 m s$^{-1}$, and an eccentricity
of 0.084$\pm$0.003. Adopting a stellar mass of 0.78$\pm$0.05 $M_{\odot}$,
we obtain a minimum mass for the companion of 13.13$\pm$0.56 $M_{\rm JUP}$, 
which is close to the planet/brown-dwarf transition, and the semimajor
axis of 1.084$\pm$0.023 AU. The nearly circular orbit despite the
large mass and intermediate orbital period makes this
companion unique among known substellar companions.

\end{abstract}



\section{Introduction}
~~During the past decade, precise Doppler surveys have observed
3000 of the closest and brightest main sequence stars
and have detected more than 250 extrasolar planets so far
(e.g. Butler et al. 2006; Udry \& Santos 2007).
The planets show a surprising diversity
in their masses, orbital periods, and eccentricities, and the statistical
distribution and correlation of these parameters now begin to serve
as the test cases for theories of planet formation and evolution
(e.g. Ida \& Lin 2004a, 2004b, 2005).

~~Stellar metallicity is one of the key parameters that controls planet
formation. It is well known that frequency of giant planets becomes higher
as stellar metallicity increases (e.g. Santos et al. 2003; Fischer
\& Valenti 2005 and references therein). Such a trend favors
a core accretion model as the main mechanism of giant planet formation,
because high metallicity likely increases the surface density at 
the midplane of the protoplaentary disk, making it easier to build 
metal cores massive enough to accrete gas envelope
(e.g. Ida \& Lin 2004b; Alibert et al. 2004).
It has also been noted that there is a possible flat tail in the 
low-metallicity regime of the frequency distribution (Santos et al. 2004;
Sozzetti et al. 2009 in press).
This might suggest the existence of a distinct  formation mechanism, disk
instability, which is not dependent on metallicity (Boss 2002).
Correlations between metallicity and orbital parameters are inconclusive. 
No significant trends have been found in period-metallicity
and eccentricity-metallicity distribution, although hints of some 
weak correlations have been pointed out (e.g. Santos et al. 2003;
Fischer \& Valenti 2005).

It has also been pointed out that in addition to the presence of 
gas giant planets, the mass of the detected planet or the
total mass of all planets in a system 
are correlated with high metallicity (e.g., Santos et al. 2003;
Fischer \& Valenti 2005), as predicted by core accretion.
Massive companions with $\gtrsim 10$ Jupiter-mass ($M_{\rm JUP}$)
could be formed by core-accretion depending on the assumed
truncation condition for gas accretion (Ida \& Lin 2004b; Alibert et al. 2004;
Mordasini et al. 2007). Or they could be formed by gravitational 
disk instability as has been suggested for the formation of brown-dwarf
companions. In fact, an upper limit of planet mass has not been well
established theoretically or observationally. 
In core accretion, the limit is regulated by a balance between gap opening
(e.g., Ida \& Lin 2004a; Crida et al. 2006), the
inflow rate due to viscous diffusion and dissipation timescales 
of protoplanetary
disks (e.g. D'Angelo et al. 2003; Dobbs-Dixon et al. 2007;
Tanigawa \& Ikoma 2007). Planet mass might also
depend on subsequent evolution of planets by mechanisms 
such as giant impacts (Baraffe et al. 2008).
In disk instability, planet mass depends primarily 
on the mass ratio between the disk and
central star. Observational properties of planets help to 
constrain the planet formation models. Currently, no significant
difference in stellar metallicity has been found for planets above and
below $\sim10M_{\rm JUP}$ (e.g., Santos et al. 2003; Fischer \& Valenti 2005).
Indeed, very high mass planets (i.e., greater than $10 M_{\rm JUP}$) are rare;
a larger sample would improve statistics for extracting information
regarding the upper limit of planet mass and correlations between planet
mass and metallicity.

The N2K consortium began precise Doppler surveys at Keck, Magellan,
and Subaru in 2004, targeting a new set of 2000 metal-rich solar-type stars 
(Fischer et al. 2005). The main purpose of the survey was to search
for short-period planets with a high-cadence observational strategy 
in order to find
prospective transiting planets. The sample was biased toward high-metallicity
stars in order to increase the detection rate of the planets.
From the collective N2K surveys, we have discovered 7 short-period 
($P \leq 5$ d) planets so far including: HD 88133 (Fischer et al. 2005), 
the transiting planet HD 149026 (Sato et al. 2005),
HD 149143 and HD 109749 (Fischer et al. 2006), 
HD 86081, HD 224693, and HD 33283 (Johnson et al. 2006).

The metallicity-biased sample is expected to contain many
long-period planets as well as short-period ones.
Therefore, we have continued to observe stars with significant radial 
velocity variations and we have also 
detected 13 intermediate-period (18--1405 d) planets: 
HD 5319 and HD 75898 (Robinson et al. 2007), 
HD 11506, HD 125612, HD 170469, HD 231701 and the transiting 
planet HD 17156 (Fischer et al. 2007), the double planet system HIP 14810
(Wright et al. 2007), HD 205732 and HD 154672 (Lopez-Morales 2008), 
HD 179079 and HD 73534 (Valenti et al 2009 in press).
These planets help to investigate correlations between properties of
planets and stellar metallicity in more detail, especially in the
high-metallicity regime.

Here, we report the discovery of a substellar companion to the G5 dwarf
star HD 16760 in a 1.3 yr nearly-circular orbit from the N2K sample.
The companion has a minimum mass of 13.13 $M_{\rm JUP}$, which is just above
the deuterium-burning threshold that is often used to distinguish 
brown dwarfs from planets. 

\section{Stellar Parameters}
HD 16760 (HIP 12638) is listed in the Hipparcos catalogue (ESA 1997)
as a G5V star with a visual magnitude $V=8.7$ and a color index
$B-V=0.715$. The revised Hipparcos parallax of $\pi=22.00\pm2.35$ mas
(van Leeuwen 2007) corresponds
to a distance of 45.5 pc and yields the absolute visual magnitude
of $M_V=5.41$. The star is probably a visual binary system having a
Hipparcos double star catalogue entry. The secondary is separated
by 14.6 arcsec which corresponds to a projected separation of about 660 AU.
A high-resolution spectroscopic analysis described
in Valenti \& Fischer (2005) derives an effective temperature
$T_{\rm eff}$ = 5629$\pm$44 K, a surface gravity $\log g$ = 4.47$\pm$0.06,
rotational velocity $v\sin i$ = 0.5$\pm$0.5 km~s$^{-1}$, 
and metallicity [Fe/H] = 0.067$\pm$0.05 dex for the star.
The bolometric luminosity is $L_{star}=0.72\pm0.43{\lsun}$ calculated
using the $M_V$ and a bolometric correction of $-$0.108 based
on Flower (1996). The radius of 0.81$\pm$0.27 $\rsun$ for the star is
derived from the Stefan-Boltzmann relation using $L_{star}$ and
$T_{\rm eff}$.

To estimate a stellar mass, we interpolated the
metallicity, effective temperature, and luminosity onto the stellar
interior model grids computed by Girardi et al. (2002). We used the
three-dimensional interpolation method described by Johnson
et al. (2007), and adopt their estimated 7\% uncertanty based
on a comparison among different stellar model grids. Using the
SME--derived stellar parameters, we estimate
$0.78 \pm 0.05 M_\odot$ for HD 16760.

As an indicator of chromospheric activity, we measured $S_{HK}$,
the core emission in the Ca II HK lines relative to the continuum,
to be 0.176 for the star. The ratio of flux from $S_{HK}$ to the
bolometric stellar flux, $\log R'_{HK}$, was derived to be
$\log R'_{HK}=-4.93$, which indicates that the star is chromospherically
inactive. The expected stellar ``jitter'', which is intrinsic variability
in radial velocity as an additional source of astrophysical noise,
was estimated to be 2 m s$^{-1}$ for the star based on the
activity and the spectral type from empirical relation by Wright (2005).
We applied the jitter to radial velocities from both Subaru and Keck
when fitting a Keplerian model in section 3.
The stellar parameters are summarized in Table 1.

\section{Radial Velocities and Orbital Solutions}
Ten radial velocity observations of HD 16760 were obtained with the
High Dispersion Spectrograph (HDS) on the 8.2 m Subaru Telescope
(Noguchi \etal 2002) from December 2004 to February 2008.
We used an iodine (I$_2$) absorption cell (Kambe \etal 2002) to provide a
fiducial wavelength reference for precise radial velocity measurements.
We adopted the setup of StdI2b for all the data, which simultaneously
covers a wavelength region of 3500--6100 ${\rm \AA}$ by a mosaic
of two CCDs. The slit width was set to 0$^{\prime\prime}$.8 for the
first 4 data (2004--2005) and 0$^{\prime\prime}$.6 for the last 6 ones
(2006--2008), giving a reciprocal resolution ($\lambda/\Delta\lambda$) of
45000 and 60000, respectively. Typical signal-to-noise ratio (S/N) was
140--200 pix$^{-1}$ with exposure time of 110--300 s depending on the
observing conditions. Radial velocity analysis for
an I$_2$-superposed stellar spectrum was carried out with a
code developed by Sato et al. (2002), which is based on a technique
by Butler \etal (1996) and Valenti \etal (1995), giving a Doppler
precision of about 4--5 \ms.

After the first 3 observing runs at Subaru, we identified significant radial
velocity variations of the star and then began follow-up observations
with the 10-m Keck telescope. Seventeen Keck radial velocity data were obtained
with the HIRES spectrograph (Vogt et al. 1994) between Jan 2006 and Jan 2009. 
We used the B5 decker ($0\farcs 86$ width with a resolution of 
about $R = 65 000$). Exposure times ranged from 150 to 240 seconds
depending on the observing conditions, resulting a consistent 
S/N of about 150. An I$_2$ cell was inserted in the 
light path to provide the wavelength solution and the instrumental 
PSF (Marcy et al. 1992; Butler et al. 1996) in our model of the observed
spectrum. The typical Doppler precision for the Keck observations is 
1.5 $\ms$. 

The Doppler measurements are listed in Table 2 along with the
time of observation, the estimated uncertainties and the origin 
of the observation (Subaru or Keck). The radial velocities were 
modeled with a Keplerian orbit using a Levenberg-Marquardt fitting 
algorithm to obtain a minimum chi-squared solution by varying 
the free parameters (orbital period, time of periastron passage, 
eccentricity, velocity amplitude and omega - the orientation of the orbit 
reference to the line of nodes). The combined radial velocities from 
Subaru and Keck are plotted in Figure 1 and the best Keplerian fit is 
overplotted as a solid line. The expected stellar jitter of 2 m s$^{-1}$ 
was added in quadrature to the velocity uncertainties when fitting 
the Keplerian orbit and are included in the velocities plotted in Figure 1. 

The best fit orbital parameters are listed
in Table 3. The uncertainty for each orbital parameter was determined 
using a bootstrap Monte Carlo approach, subtracting the theoretical fit, 
scrambling the residuals, adding the theoretical fit back to the residuals 
and then refitting. The radial velocities are best fit by a Keplerian model
with a period $P=466.47\pm0.35$ d, a velocity semiamplitude
$K_1=407.71\pm0.84$ \ms, and an eccentricity $e=0.084\pm0.003$.
An offset of 37 m s$^{-1}$ was applied to the Subaru data
in order to minimize reduce chi-squared ($\sqrt{\chi_{\nu}^2}$) when fitting a
Keplerian model to the combined Subaru and Keck velocities.
The rms scatter of the residuals to the Keplerian fit is 4.3 m s$^{-1}$,
and the reduce chi-squared is $\sqrt{\chi_{\nu}^2}=1.3$.
We found no significant additional periodicity in the residuals.
Adopting a stellar mass of
0.78$\pm$0.05 \msun, we obtained for the companion a minimum mass
$M_p\sin i=13.13\pm0.56$ $M_{\rm JUP}$, which is close to the
planet/brown-dwarf transition, and a semimajor axis $a=1.084\pm0.023$ AU.

If we assume the orbit is randomly oriented, there is a 1.4\%
chance that the true mass exceeds 80 $M_{\rm JUP}$ ($i<9.5^{\circ}$),
the boundary between the brown dwarf and stellar mass regime.
In this case, the projected semi-major axis is less than 2.1 mas
based on the $a_1\sin i$ and the distance to the star,
which is below the measurement error of Hipparcos.
We measured a small projected rotational velocity for HD 16760
(\vsini$=$0.5 \ks). This may suggest that a near pole-on viewing
angle and thus high mass of the companion is possible if the orbital
plane is co-planar to the stellar equator.
High-precision spaceborn astrometric observations such as using Hubble Space
Telescope, which can achieve $\sim0.3$ mas precision (e.g., Bean \etal 2007),
are highly encouraged to set a stringent constraint on the mass of the
companion.

\section{Discussion and Summary}
We have reported the discovery of a substellar companion around
the G5 dwarf star HD 16760 in a 1.3 yr-orbit from precise Doppler
measurements at Subaru and Keck.
Although the N2K survey originally targeted short-period planets
with a few successive observations, based on the Fischer and Valenti 
metallicity correlation, we expect that 15\% of the target stars harbor 
gas giant planets. The metallicity-biased sample is expected to 
contain many long-period planets as well as additional short-period 
ones. 

HD 16760 b has a minimum mass of about 13 $M_{\rm JUP}$, which is close
to the border between planet and brown dwarf regimes. By definition,
this is a brown dwarf because the mass is above limit for deuterium
buring ($13M_{\rm JUP}$) and in this context it is considered to be part of a
low mass tail of the brown-dwarf distribution. However, such a companion
is often called a ``super-planet'' because it could be a high mass
tail of the planet distribution too.
Core-accretion models of planet formation can predict super-massive
planets larger than 10 $M_{\rm JUP}$ or even up to 20--25 $M_{\rm JUP}$
under metal-rich environment, depending on the assumed truncation condition
for gas accretion (Ida \& Lin 2004b; Alibert et al. 2004;
Mordasini et al. 2007). In fact, the
upper limit of planet mass has not been well established theoretically,
and it should be set based on the observational properties of planets.

Currently, 7 companions with minimum mass of 13--25 $M_{\rm JUP}$ have
been detected within semimajor axis of 5 AU around solar-type stars
(FGK dwarfs with $0.7\le M/M_{\odot}<1.6$) by precise Doppler measurements
including a transiting one CoRoT-Exo-3 b for which accurate mass of
21.7 $M_{\rm JUP}$ was obtained (Deleuil et al. 2008).
The number of such companions is still small and their
statistical properties have not yet been established.
In Figure \ref{fig3}, the planet eccentricity is plotted against planet mass.
As pointed out by Marcy et al. (2005), the more massive planets
($\gtrsim 5M_{\rm JUP}$) have systematically higher eccentricities than
lower-mass planets. Most companions with $>8M_{\rm JUP}$
(other than HD 16760 b and HD 168443 c\footnote{HD 168443 has two planets;
one is 8 $M_{\rm JUP}$ (b) and the other is 18 $M_{\rm JUP}$ (c)}) have
eccentricities
larger than about 0.3 except for planets in short-period orbits where
tidal circularization could be effective.

Since massive companions have the largest inertial resistance to
perturbations that drive them out of their initial orbits,
they should retain their initial orbits. Therefore, 
massive planets with high eccentricity are difficult
to explain within the framework of standard core accretion model.
This suggests a distinct evolutionary scenario that could result in
eccentric orbits for massive companions, such as disk instability 
(e.g. Boss 1998), giant impacts between massive planets following 
core accretion (Baraffe et al. 2008), or orbital evolution by 
Kozai mechanism for binary stars (e.g. Wu \& Murray 2003).
The fact that HD 16760 is probably in a binary system but the companion
has a small eccentricity is interesting from this view point of
orbital evolution with respect to the Kozai mechanism.

The metallicity of the host stars may be one of the keys to distinguishing
between planet formation models; disk instability is not dependent on
metallicity (Boss 2002), while accretion should be more efficient 
in high metallicity disks. We do not observe a correlation between
metallicity and eccentricity for the most massive companions at this stage.
However, there are still only a small number of these objects.

In summary, HD 16760 b is unique among the known substellar companions 
because it resides in a nearly-circular orbit with an intermediate
orbital period and a mass that is greater than 13 times the mass of 
Jupiter. Detection of additional objects in this mass regime 
could provide deeper insight for theoretical models.

\acknowledgements
We thank Akito Tajitsu and Tae-Soo Pyo for their expertise and support
of the Subaru HDS observations. We gratefully acknowledge the dedication
and support of the Keck Observatory staff, in particular Grant Hill for
support with HIRES. 
B.S. is supported by MEXT's program "Promotion of Environmental
Improvement for Independence of Young Researchers" under the Special
Coordination Funds for Promoting Science and Technology. DF gratefully 
acknowledges support from NASA grant NNG05G164G.
This research has made use of the Simbad database,
operated at CDS, Strasbourg, France. The authors extend thanks to those of
native Hawaiian ancestry on whose sacred mountain of Mauna Kea we are
privileged to be guests. Without their generous hospitality, the Subaru and
Keck observations presented herein would not have been possible.

\clearpage



\begin{figure}
\plotone{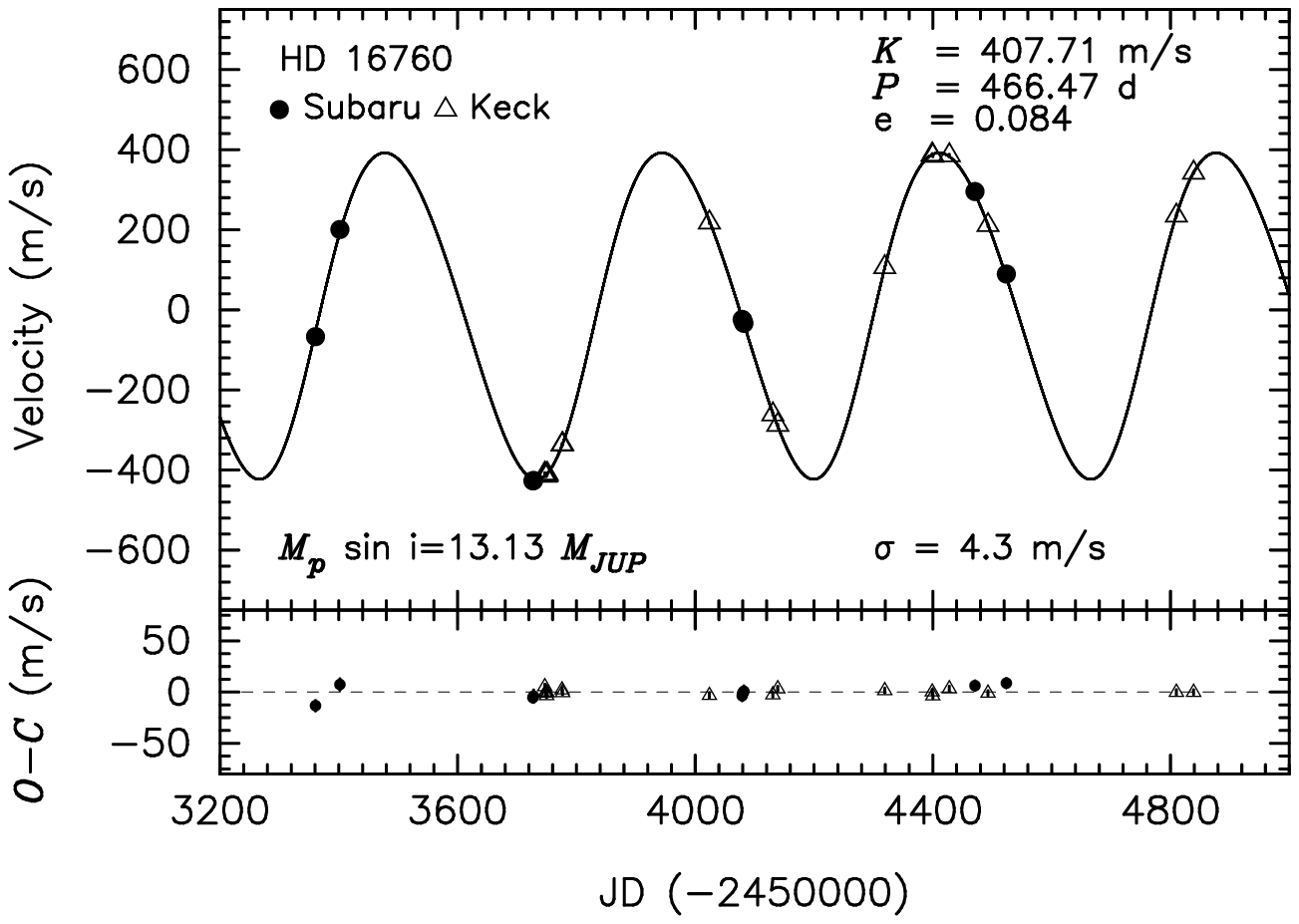}
\caption{Radial velocities for HD~16760 from
Subaru (filled circles) and Keck (open triangles).\label{fig1}}
\end{figure}

\clearpage
\begin{figure}
\plotone{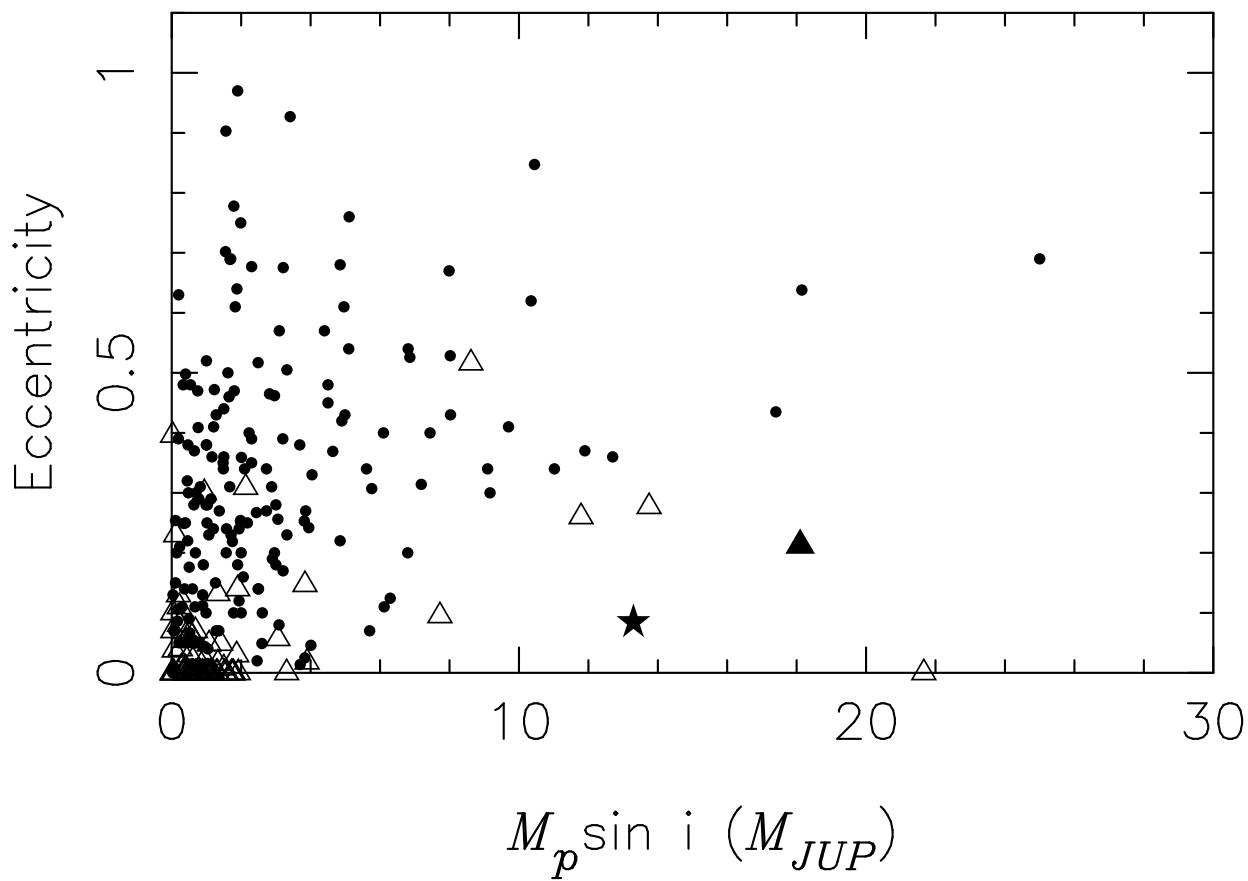}
\caption{Eccentricity plotted against minimum mass of the
planet. HD 16760 b is plotted with {\it star}. HD 168443 c
is plotted with filled triangle, in which system there is another
planet with 8 $M_{J}$ (HD 168443 b). CoRoT-Exo-3 b, a transiting
brown dwarf/super-planet, has a mass of 21.7 $M_{J}$ without $\sin i$
ambiguity in a 4.26-d orbit. Planets with $a<0.1$AU
({\it open triangle}) may be affected by tidal effect. Only planets
around FGK dwarfs are plotted in the figure (data from
http://exoplanet.eu).\label{fig3}}
\end{figure}

\clearpage

\begin{deluxetable}{lll}
\tablenum{1}
\tablecaption{Stellar Parameters for HD 16760}\label{tbl-1}
\tablewidth{0pt}
\tablehead{\colhead{Parameter}  & \colhead{Value} \\
} 
\startdata
$V$                & 8.7          \\
$M_V$              & 5.41          \\
$B-V$              & 0.715      \\
$B.C.$              & $-$0.108      \\
Spectral type      & G5V        \\
Parallax (mas)     & 22.00 (2.35)    \\
$T_{\rm eff}$ (K)   & 5629 (44)   \\
\logg              & 4.47 (0.06)   \\
${\rm [Fe/H]}$     & +0.067 (0.05)   \\
\vsini (\ks)       & 0.5 (0.5)     \\
$M_{STAR}$ (\msun) & 0.78 (0.05)    \\
$R_{STAR}$ (\rsun) & 0.81 (0.27)    \\
$L_{STAR}$ (\lsun) & 0.72 (0.43)    \\
$S_{HK}$          &  0.176 \\
$\log R'_{HK}$    &  $-$4.93 \\
\enddata                         
\end{deluxetable}                           

\clearpage

\begin{table}
\tablenum{2}
\caption{Radial Velocities for HD~16760}\label{tbl-2}
\begin{tabular}{rrrc}
\tableline
\tableline
       JD  &   Radial Velocity    & Uncertainties  & Observatory \\
 $-$2450000  &  (\ms)  &   (\ms)        &             \\
\tableline
3360.89840 & $-$67.01 & 4.66 & Subaru\\
3401.87152 & 200.38 & 5.28 & Subaru\\
3726.87567 & $-$427.26 & 4.15 & Subaru\\
3727.85200 & $-$426.31 & 5.70 & Subaru\\
3746.81122 & $-$409.23 & 1.28 & Keck\\
3747.87456 & $-$414.10 & 1.24 & Keck\\
3748.78525 & $-$411.79 & 1.20 & Keck\\
3749.78054 & $-$413.81 & 1.16 & Keck\\
3750.79189 & $-$409.55 & 1.15 & Keck\\
3775.77328 & $-$336.64 & 0.95 & Keck\\
3776.81038 & $-$335.11 & 1.25 & Keck\\
4023.95588 & 217.93 & 1.50 & Keck\\
4078.93737 & $-$24.05 & 4.52 & Subaru\\
4079.92984 & $-$26.00 & 4.99 & Subaru\\
4080.99173 & $-$30.70 & 4.60 & Subaru\\
4081.88938 & $-$33.80 & 4.83 & Subaru\\
4130.74043 & $-$261.06 & 1.10 & Keck\\
4138.79826 & $-$287.38 & 1.11 & Keck\\
4319.10806 & 106.52 & 1.14 & Keck\\
4398.95389 & 387.91 & 1.13 & Keck\\
4399.88492 & 384.74 & 1.36 & Keck\\
4427.84750 & 385.80 & 1.43 & Keck\\
4470.89963 & 295.34 & 4.17 & Subaru\\
4492.75215 & 211.30 & 1.10 & Keck\\
4523.74114 & 89.43 & 4.05 & Subaru\\
4809.84453 & 235.27 & 1.52 & Keck\\
4838.85261 & 342.52 & 1.02 & Keck\\
\end{tabular} 
\end{table}

\clearpage

\begin{deluxetable}{ll}
\tablenum{3}
\tablecaption{Orbital Solution for HD 16760b}\label{tbl-3}
\tablewidth{0pt}
\tablehead{\colhead{Parameter}  & \colhead{Value} \\
} 
\startdata
$P$ (days)               &  466.47 (0.35)   \\
$T_{\rm p}$ (JD)         &  2453337.0 (2.4) \\
Eccentricity             &  0.084 (0.003)    \\
$\omega$ (deg)           &  242.9 (1.9)     \\
$K_1$ (\ms)              &  407.71 (0.84)   \\
$a$ (AU)                 &  1.084 (0.023)     \\
$a_1 \sin i$ (10$^{-3}$AU) &  17.420 (0.042)      \\
$f_1$(m) (10$^{-6}$\msun)  &  3.241 (0.021)   \\
$M_p\sin i$ (\mjup)        &  13.13 (0.56)    \\
${\rm N_{obs}}$ (Subaru)  &   10          \\
${\rm N_{obs}}$ (Keck)    &   17          \\
RMS (\ms)                &   4.3       \\
Reduced $\sqrt{\chi_{\nu}^2}$  &   1.3       \\
\enddata                        
\end{deluxetable}                          

\clearpage

\end{document}